\documentclass[prb,aps,amssymb,twocolumn,showpacs,floatfix]{revtex4}
\usepackage{graphics}
\usepackage{epsfig} 
\usepackage[dvips]{color}
\usepackage{dcolumn}
\usepackage{amsmath}
\usepackage{color}
\begin{document}
\newcommand\dd{{\operatorname{d}}}
\newcommand\sgn{{\operatorname{sgn}}}
\def\Eq#1{{Eq.~(\ref{#1})}}
\def\Ref#1{(\ref{#1})}
\newcommand\e{{\mathrm e}}
\newcommand\cum[1]{  {\Bigl< \!\! \Bigl< {#1} \Bigr>\!\!\Bigr>}}
\newcommand\vf{v_{_\text{F}}}
\newcommand\pf{p_{_\text{F}}}
\newcommand\ef{{\varepsilon} _{\text{\sc f}}}
\newcommand\zf{z_{_\text{F}}}
\newcommand\zfi[1]{{z_{_\text{F}}}_{#1}}
\newcommand\av[1]{\left<{#1}\right>}
\def\det{{\mathrm{det}}}
\def\Tr{{\mathrm{Tr}}}
\def\Li{{\mathrm{Li}}}
\def\tr{{\mathrm{tr}}}
\def\im{{\mathrm{Im}}}
\def\Texp{{\mathrm{Texp}\!\!\!\int}}
\def\antiTexp{{\mathrm{\tilde{T}exp}}\!\!\!\int}
\title{Cold bosons in the Landauer setup} 

\author{D. B. Gutman$^{1,2}$, Yuval Gefen$^3$, and A. D. Mirlin$^{2,4,5}$}
\affiliation{
\mbox{$^1$Department of Physics, Bar Ilan University, Ramat Gan 52900,
Israel }\\
\mbox{$^2$Institut f\"ur Nanotechnologie, Karlsruhe Institute of Technology, 
 76021 Karlsruhe, Germany}\\
\mbox{$^3$Dept. of Condensed Matter Physics, Weizmann Institute of
  Science, Rehovot 76100, Israel}\\
$^4$Institut f\"ur Theorie der kondensierten Materie and DFG
  Center for Functional Nanostructures,\\ 
Karlsruhe Institute of Technology, 76128 Karlsruhe, Germany\\
\mbox{$^5$Petersburg Nuclear Physics Institute, 188300 St.~Petersburg, Russia}
}

\date{\today}

\begin{abstract}
We consider one dimensional potential trap that connects
two reservoirs containing  cold Bose atoms.
The thermal current and  single-particle bosonic Green 
functions are calculated under non-equilibrium conditions. 
The bosonic statistics leads to  Luttinger liquid state 
with non-linear spectrum of collective modes.
This results in suppression of thermal 
current at low temperatures
and affects  the single-particle Green functions.     
\end{abstract}

\maketitle

\section{Introduction}
\label{s1}

Systems of ultracold  Bose  gases have recently attracted  a great
deal of experimental  \cite{Kinoshita,Hofferbeth,Amerongen,Paredes} 
and theoretical attention, see 
Refs. \onlinecite{Zwerger,Giamarchi} for reviews.  
A high control over experimental  conditions, including geometry,
density, and interaction strength, as well as absence of uncontrolled
disorder allow one to explore new aspects of many-body physics. 
Among experimental achievements, 
the  coherence of non-equilibrium  Bose condensate was studied based
on the interference measurement  
\cite{Hofferberth_2008},  correlations of density fluctuations  were
measured in Ref.\onlinecite{ Esteve}, and  
the distribution of the bosons over momenta was explored in Ref.
\onlinecite{Amerongen,Paredes,Richard}. 

Unlike the classic example of a bulk Bose fluid
($^4$He), atomic gases are usually realized in optical traps or in
atom chips where magnetic and electric fields confine the system to a
geometry with strong asymmetry with respect
to three-dimensional  rotations. In many experimental situations,  
one deals with arrays of quasi-1D systems\cite{Hofferbeth,Richard,Davidson}.   
These geometrical restrictions  strongly influence the dynamics as they 
lower  the effective dimensionality of the system.
Indeed, in  three dimensions a Bose system undergoes a famous
Bose-Einstein condensation, 
its thermodynamic properties are  well accounted by the mean free 
theory \cite{Pitaevskii_book}, while its hydrodynamics is governed by the
Gross-Pitaevskii equation. On the other hand,  
in two dimensions and especially in one dimension 
fluctuations of the order parameter destroy 
the long range order, necessitating a more microscopic treatment.  
In this work we focus on the impact these effects have  on transport 
properties of one-dimensional bosonic systems.

\begin{figure}
\includegraphics[width=\columnwidth,angle=0]{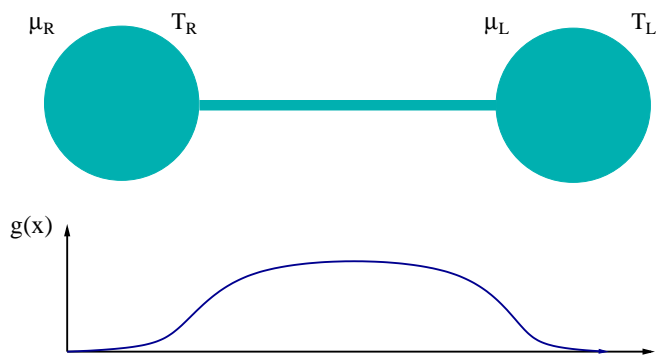}
\caption{Two reservoirs with cold atoms are connected by
  one-dimensional trap.   
The temperatures and chemical potentials in the reservoirs are assumed
to be different  
(upper panel); interaction strength $g(x)$ as a function of coordinate
(lower panel).} 
\label{setup}
\end{figure}

As is well known, a clean  
one-dimensional (1D) system forms a strongly correlated ground state,
so-called Luttinger liquid (LL).
Though this description holds for both, fermionic and bosonic systems, 
the  bosonic character begets new properties in Luttinger liquid state. 
To explore these features, we consider 
Landauer type setup shown in Fig.\ref{setup}, 
where bosons are trapped in the system that consists of two reservoirs 
connected by a one dimensional ``wire''.

Far-from-equilibrium realizations of Landauer setup have 
been recently studied in the framework of correlated 1D electronic (or,
more generally, fermionic) 
systems. The tunneling spectroscopy of the
non-equilibrium carbon nanotube have been measured in
Ref.~\onlinecite{tunnel-spectroscopy}. 
The thermal current in the edge state of QHE was studied in 
Ref.~\onlinecite{tunnel-spectroscopy-qhe}.
On the theory side, 
one can distinguish several types of non-equilibrium setups,
ranging from partially \cite{gutman08,gutman09} 
to fully non-equilibrium situations \cite{GGM_long2010}.
In the partial-equilibrium setup, electrons 
coming from different reservoirs have different values of
chemical potential and temperature. 
In the case of full non-equilibrium, electrons in the reservoirs
are characterized by an arbitrary single-particle density matrix. 
Remarkably, correlation functions of the interacting many-body problem
can be calculated exactly even in the latter case, and  can be cast in
terms of Fredholm determinants.  
 
In this paper, we address analogous questions 
in the context of bosonic system. 
Though at this moment we are not aware of any direct experimental 
realization of Landauer-type setup for bosons, the idea 
seems experimentally feasible.  
Indeed, the confinement of bosons to 1D optical wire has been 
accomplished in Ref.\onlinecite{Esteve}. 
Since the case of partial equilibrium seems 
to be more natural from the point of view of experimental realization, 
we focus on it in the current work. 
We assume that the interaction between the atoms is of a hard core type.
Inside the reservoirs it plays  little role, and  we approximate the atoms 
there as an ideal Bose gas. 
Inside the 1D ``wire'' connecting the reservoirs
the hard core repulsion can not be neglected.
We  thus describe the system by LL with spatially varying interaction
parameter $g(x)$, see Fig.\ref{setup}.  
One of the key features distinguishing this bosonic setup from its
fermionic counterpart is the absence of  Fermi surface.  In other words, the
excitation spectrum of particles in the non-interacting regions of the
system is quadratic.  Another interesting realization of  
excitations with quadratic spectrum  are transverse spin waves in a 
ferromagnetic Bose gas\cite{Zvonarev}.

The non-linearity of the excitation spectrum is known to have  a significant  
impact on the properties of the LL. In the fermionic case it leads  to
a number of interesting and important effects, see
Refs.~\onlinecite{Glazman2009,Glazman2010,Glazman-RMP}  
for reviews. However, for most characteristics of low-temperature dynamics
of correlated electronic systems such  non-linearity  can be discarded. 
This  is done indeed in the case of  the standard LL model.
In the present situation, the spectrum for weak interaction is not linear
even in the leading approximation. This should be accounted for in
the corresponding theory and may be expected to profoundly
affect the results of our analysis.

To deal with a non-linear dispersion of the spectrum 
in the bosonic case, we use the so-called harmonic approximation
\cite{Haldane}.  
Remarkably, this  approach  accurately describes the system in both the 
``quasi-condensate'' (weakly interacting Bose gas) and  the LL
(sufficiently strong repulsive interaction) regimes.
Within this framework, we study the thermal current and  
single-particle Green functions. The latter contain  
information about the density of states and the 
distribution function of bosons, as well as 
about information about phase-coherence correlations that are 
probed in interference experiments
\cite{Hofferbeth}.   

\section{Bosonization of bosons}
\label{s2}

We begin with  the Hamiltonian 
\begin{equation}
\label{Hamiltonian_1}
H=H_0+H_{\rm int}\,,
\end{equation}
that consists of the free part (we set $\hbar=1$),  
\begin{equation}
H_0=-\frac{1}{2m}
\int dx \hat{\Psi}^\dagger \partial_x^2\hat{\Psi}\,,  
\end{equation}
and the  interaction,
\begin{equation}
H_{\rm int}=\int dxdx' \hat{\rho}(x)V(x,x')\hat{\rho}(x')\,.
\end{equation}
Here we define the density of the bosonic field 
$\rho(x)=\Psi^\dagger(x)\Psi(x)$;
the bosonic field $\Psi$ satisfies canonical commutation relations,
\begin{equation}
[\hat{\Psi}(x),\hat{\Psi}^\dagger(x')]=\delta(x-x')\,.
\end{equation}

To analyze the problem we use the hydrodynamic approach \cite{Haldane}, 
similar to the bosonization for fermionic systems.
The term ``bosonization'' in the present context is,
perhaps, not optimal since the original system is
bosonic to begin with. What actually happens is a transformation from the
original bosonic fields $\Psi_B$, $\Psi_B^\dagger$ 
to new collective degrees of freedom
described by bosonic fields $\phi, \theta$.  
The original field operator is expressed in term of the new fields as
\cite{Popov},\cite{Haldane_note} 
\begin{equation}
\label{bose_operator}
\hat{\Psi}_B(x)=\sqrt{\rho+\hat{\Pi}(x)}e^{i\hat{\theta}(x)}\,. 
\end{equation}
Here the field $\hat{\phi}(x)$ is related to the smeared density 
$\hat{\rho}(x) = \rho +
\hat{\Pi}(x)$ (where $\rho$ is the average density) via $\hat{\Pi}(x)
=-\partial_x\hat{\phi}(x)/\pi$. 
The collective bosonic fields ($\hat{\phi}, \hat{\theta}$) satisfy the commutation relation
\begin{equation}
[\hat{\phi}(x),\hat{\theta}(x')]=\frac{i\pi}{2}\rm{sgn}(x-x')\,\,.
\end{equation}
The substitution of  Eq.~(\ref{bose_operator}) into 
the Hamiltonian (\ref{Hamiltonian_1}) leads 
to a hydrodynamic description of 1D bosons \cite{Pitaevskii_book}.
This is a non-linear theory that  can be considerably simplified 
by using the harmonic approximation. 
Expanding the Hamiltonian in the fields $\theta,\phi$, one keeps
terms only up to the quadratic level, which yields 
\begin{equation}
\label{bos_Hamiltonian1}
H_0=\frac{1}{2m}\int dx
\bigg[\frac{1}{4\rho}(\partial_x\hat{\rho})^2+
\rho(\partial_x\hat{\theta})^2\bigg]\,. 
\end{equation}
The harmonic approximation allows us to account for
the non-linear (in the present case quadratic) spectrum of the low energy 
sound mode in the collective theory.
Its validity is restricted to 
the low-energy (large-density) regime, $T\ll {\rm max}\{\Lambda,\rho g\}$. 
Here $\Lambda=\rho^2/m$ is the bosonic counterpart of Fermi energy, and
$g$ is the interaction strength introduced below.  

Clearly, the harmonic approximation is not an exact theory. 
By neglecting the interaction between 
low-energy modes (represented by terms of higher order in $\hat{\rho}$
and $\hat{\theta}$ in the Hamiltonian), 
one discards relaxation precesses that are 
important, in particular, for thermal equilibration, drag, and
thermoelectric 
effect \cite{Glazman2010,Glazman-RMP,Khodas2006,Khodas2007,Aristov,Gangardt-Kamenev} .  We will assume
that our ``wire'' is not too long, so that 
neglecting these processes is justified.
In that situation the harmonic approximation is
sufficient to describe thermal transport 
and tunneling spectroscopy in the system.

In terms of the collective bosonic 
fields $\theta$ and $\phi$ we obtain
\begin{eqnarray}&&
\label{bos_Hamiltonian2}
H_0=\int dx\bigg[\frac{1}{8\pi^2 m\rho(x)}\left(\partial_x^2\phi\right)^2+
\frac{\rho(x)}{2m}\left(\partial_x\hat{\theta}\right)^2\bigg]\, ,\nonumber  \\&&
H_{\rm int}= \int dx \frac{g(x)}{\pi^2}\left(\partial_x\hat{\phi}\right)^2\,.
\end{eqnarray}
Here we model the interaction with a short range potential,
\begin{equation}
\label{local-int}
V(x,x')=g(x)\delta(x-x')\,.
\end{equation}
Let us note that Eq.~(\ref{bos_Hamiltonian2}) in fact corresponds to
the microscopic model (\ref{local-int}) in the limit of weak
interaction only. In the  local interaction model (\ref{local-int}), 
the large-$g$ limit yields the Tonks-Girardeau gas\cite{Girardeau} 
that can be mapped onto a free-fermion model 
(characterized by the LL parameter $K=1$).
However, it makes sense to consider $g$ in
Eq.~(\ref{bos_Hamiltonian2}) as a phenomenological parameter
of underlying LL model, which allows us to go beyond the Tonks-Girardeau limit. 
On a microscopic level, this corresponds to  the replacement of the 
delta-like repulsion (\ref{local-int}) by a finite-range hard-core
interaction.

Equation (\ref{bos_Hamiltonian2}) is the 
Hamiltonian of a 1D LL for  interacting bosonic system
in the harmonic approximation \cite{Giamarchi}.
As one clearly observes, the Hamiltonian (\ref{bos_Hamiltonian2}) contains
a fourth-order spatial derivative [the term
$(\partial_x^2\hat{\phi})^2$], at odds with the standard 
fermionic LL model that contains only a second spatial derivative
[$(\partial_x\hat{\phi})^2$]. 
For large values of the interaction constant,   
$\rho g \gg  \rho^2/2m$, 
the fourth-derivative term is relatively small, leaving us with the standard
LL  with the interaction parameter  $K^2=\pi^2\rho/2mg$.
When the interaction is weakened, $\rho g \ll  \rho^2/2m$, the
standard (linear-spectrum) LL description is valid for  the lowest
temperatures only, $T\ll \rho g$. The LL parameter becomes larger for
weaker interaction and tends to infinity
in the limit of free bosons but the region of applicability of the
standard LL theory vanishes in this limit.

In the general case, one should use the full theory (rather than the
standard LL theory). The corresponding spectrum of bosonic excitations
$\omega(q)$ is non-linear\cite{Chalker} , interpolating between   quadratic
($\omega=q^2/2m$ for non-interacting bosons) and linear  
($\omega=uq$ for  strongly interacting bosons; $u$ being the sound velocity). 
Finally, we mention  that the  Hamiltonian (\ref{bos_Hamiltonian2}) depends on 
the mean bosonic density, $\rho(x)$.  
To find the profile of the bosonic density, one needs to go beyond  the LL 
description and solve the non-linear hydrodynamic equations, see
Appendix \ref{non_linear}. 
 
To deal with the non-equilibrium conditions, we use the Keldysh formalism.
The fields on the upper and lower branches are labeled by $+$ 
and $-$ respectively.
It is convenient to perform a rotation in Keldysh space
\begin{eqnarray}&&
\phi,\bar{\phi}=(\phi_+\pm \phi_-)/\sqrt{2}\,,\\&&
\theta,\bar{\theta}=(\theta_+\pm\theta_-)/\sqrt{2}\,.
\end{eqnarray}
where we will refer to $(\phi,\theta)$ as classical 
and to $(\bar{\phi},\bar{\theta})$ as quantum components\cite{Kamenev}.
We then find that the 
system is described by the following action
\begin{eqnarray}
\label{a4}
S=\frac{1}{2}\Phi^T D^{-1}\Phi\,, 
\end{eqnarray}
where we have defined the vector 
$\Phi=(\phi,\theta,\bar{\phi},\bar{\theta})$ \,.
The inverse propagator has a standard structure in the Keldysh space 
\begin{eqnarray}
{D^{-1}}=\begin{pmatrix}
0 & (D^{-1})^r\\
(D^{-1})^a & (D^{-1})^K
\end{pmatrix},
\label{Polarization_operator}
\end{eqnarray}
where each component is a matrix  in the  $\theta,\phi$ space.
The inverse retarded propagator is given by
\begin{eqnarray}
(D^{-1})^r=\begin{pmatrix}
\hat{K}^{-1}  & -\frac{i\omega_+}{2\pi}\partial_x\\
-\frac{i\omega_+}{2\pi}\partial_x & -\frac{1}{2m}\partial_x\rho\partial_x
\end{pmatrix}, 
\label{Polarization_operator2}
\end{eqnarray}
with
\begin{eqnarray}
\hat{K}^{-1}=\frac{\partial_x}{8\pi^2m}\bigg[
\frac{2\rho_x}{\rho^2}\partial_x-
\frac{\rho_x^2}{\rho^3}+
\partial_x\frac{1}{\rho}\partial_x-8mg\bigg]\partial_x 
\nonumber 
\end{eqnarray}
and  $\omega_+=\omega+i0$.
The advance component is related to the retarded one 
by complex conjugation,   
$(D^{-1})^a(\omega)=\left[(D^{-1})^r(\omega)\right]^*$.
The Keldysh component $(D^{-1})^K$ carries information about the
distribution functions in two reservoirs: right-moving modes coming
from the left reservoir have the temperature $T_R$, while the
left-moving modes coming from the right reservoir are characterized 
by the temperature $T_L$. 

To simplify the analysis from now on we consider the large density limit, 
approximating the mean bosonic density $\rho(x)$ by a constant.
Variation of the action with respect to the classical components of
the fields $\theta$ and $\phi$ yields the saddle point equations 
\begin{eqnarray}&&
\left(-\frac{1}{8\pi^2m\rho}\partial_x^4+\partial_x \frac{g(x)}{\pi^2}\partial_x\right)\bar{\phi}_\omega+\frac{i\omega}{2\pi}\partial_x\bar{\theta}_\omega=0\,, \nonumber \\&&
\frac{i\omega}{2\pi}\partial_x\bar{\phi}_\omega+\frac{\rho_0}{2m}\partial_x^2\bar{\theta}_\omega=0\,.
\end{eqnarray}
These two equations can be conveniently combined into the wave equation
\begin{equation}
\label{equation_J}
\bigg[\omega^2+\partial_x\left(\frac{2\rho g(x)}{m}-\frac{1}{4m^2}
\partial_x^2\right)\partial_x \bigg]J(\omega,x)=0\,,
\end{equation}
where we have introduced $J = \pi^{-1}\partial_x\bar{\theta}$. 
Equation (\ref{equation_J})  
describes propagation of plasmons inside the wire with dispersion that varies
in space (as a result of the variation of $g(x)$). 
Due to these variation, a plasmon may experience scattering but
the total number of plasmons  is conserved.
To make this conservation explicit, we multiply Eq.~(\ref{equation_J})
by $J^*$ from the left and subtract a complex-conjugated equation.
This yields the plasmons  conservation law  
\begin{equation}
\partial_t Q+\partial_x {\cal J}=0\,
\end{equation}
which has a form of the continuity equation that states that the charge 
\begin{equation}
Q=J\partial_t J^*-J^*\partial_t J
\end{equation}
is carried by the current 
\begin{eqnarray}&&
{\cal J}=\frac{2\rho g}{m}\left(J^*J_x-JJ^*_x  \right) \\&&-\frac{1}{4m^2}
\left(J^*J_{xxx}-J^*_xJ_{xx}-JJ^*_{xxx}+J_xJ^*_{xx} \right)\,. \nonumber
\end{eqnarray}

\begin{figure}
\includegraphics[width=0.8\columnwidth,angle=0]{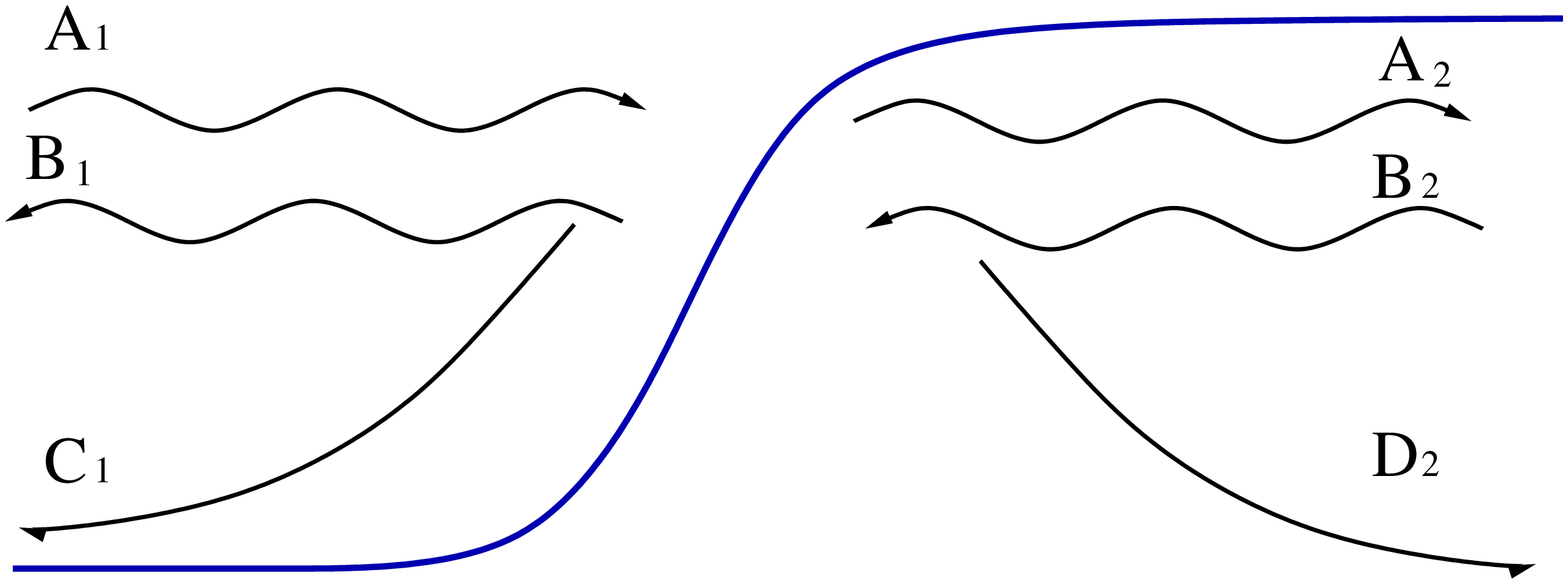}
\caption{The boundary between the regions with different interaction constants.
Propagating and decaying waves on both sides of the boundary are shown.}
\label{waves}
\end{figure}

In a region with a constant interaction  strength $g$
Eq.~(\ref{equation_J}) yields the 
Bogolubov's excitation spectrum for the acoustic phonons,  
\begin{eqnarray}&&
\omega_q^2=\frac{2\rho g}{m}q^2+\frac{q^4}{4m^2}\,. 
\end{eqnarray}
This dispersion relation has four solutions, 
resulting in  oscillating   and exponentially 
decaying (growing) waves (see Fig.~\ref{waves}),
\begin{equation}
J_{\omega}(x)= A e^{iqx}+ B e^{-iqx}+ C e^{px}+ D e^{-px}\,,
\end{equation}
where
\begin{eqnarray}&&
q=\sqrt{2}m\sqrt{-g+\sqrt{g^2+\omega^2/m^2}}, \label{q}\\&&
p=\sqrt{2}m\sqrt{g+\sqrt{g^2+\omega^2/m^2}}.   \label{p}
\end{eqnarray}

\begin{figure}
\includegraphics[width=0.9\columnwidth,angle=0]{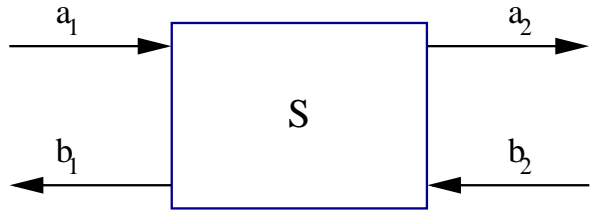}
\caption{The scattering of plasmons at the boundary between regions 
with different values of interaction can be described by scattering matrix $S$.}
\label{fig2}
\end{figure}

We consider now the situation when the interaction strength changes
from one value to another in a boundary region, see Fig.~\ref{waves}.
If we consider the solution not too close to the boundary, the
exponentially decaying components can be neglected, and the
propagating waves are related via a scattering matrix.
To take into account the different velocities of propagation, we define 
 $a=\sqrt{u}A$, $b=\sqrt{u}B$, where  $u_q=\frac{\partial
   \omega_q}{\partial q}$ is a sound velocity in corresponding
 region. It is easy to verify that coefficients  $a$ and $b$ in
 different regions  (see Fig. \ref{fig2}) are related 
\begin{eqnarray}
\left(\!\!\begin{array}{l}
a_2
\\
\\
b_1 
\end{array}
\right)=S \left(\!\!\begin{array}{l}
a_1
\\
\\
b_2
\end{array}
\right)\,
\end{eqnarray}
through a unitary scattering matrix,
\begin{eqnarray}
S=\begin{pmatrix}
t & r\\
r' & t'
\end{pmatrix} ,
\label{Scattering_matrix}
\end{eqnarray} 
with $|t| = |t'|$ and $|r|= |r'|$. 

The transmission coefficient has to be calculated for  a particular
realization $g(x)$ of the interaction in the boundary region.  
In the limiting case of an adiabatic barrier, when 
the interaction changes smoothly on the scale of the plasmon wave length, 
one finds the ideal transmission,  $|t|=1$ and $r=0$. 
We focus on the opposite case of a sharp-step barrier, with
interaction constant $g_1$ to the left of the boundary and $g_2$ to
the right.
To find the transmission and reflection amplitude of such a barrier,
we derive matching conditions for the amplitudes at the boundary. For
this purpose, we
integrate Eq.~(\ref{equation_J}) over a small region around the
boundary. This leads to a requirement that $J$, $J'$, $J''$, and
$2\rho g J' - J'''/4m$ are continuous at the boundary [which  implies
that $J'''$ has a jump equal to $8 m \rho J' (g_2-g_1)$].  
These four conditions allow us to find the amplitudes
$A_2$, $B_1$ of the outgoing waves and $C_1$, $D_2$ of the decaying waves for
given amplitudes  $A_1$ and $B_2$ of the incoming waves, and thus to
establish the scattering matrix. 
We obtain the transmission
\begin{eqnarray}&&
\label{transmission_amplitude}
t=-\frac{2}{Z}\sqrt{q_1q_2\frac{(q_1+ip_1)(q_2+ip_2)}{(q_1-ip_1)(q_2-ip_2)}}
\\&&
\times
[8\pi m\rho p_2(g_1-g_2)-(p_1+p_2)(q_1^2+p_2^2)]\,, \nonumber 
\end{eqnarray}
and the reflection amplitude
\begin{eqnarray}&&
r=\frac{1}{Z}\frac{q_2+ip_2}{q_2-ip_2}\bigg[
-(p_1+p_2)(q_1+ip_2)(q_2-ip_1)(q_1-q_2)\nonumber \\&&
+8\pi m\rho (p_1q_1+p_2q_2)(g_2-g_1)
\bigg]\,,
\end{eqnarray}
where we have introduced the notation
\begin{eqnarray}&&
Z=(p_1+p_2)(q_1+ip_2)(q_2+ip_1)(q_1+q_2)
\nonumber \\&&
+8\pi m\rho(p_1q_1-p_2q_2)(g_2-g_1).
\end{eqnarray}
For our model with non-interacting reservoirs, we now consider 
the case of the scattering between interacting and non interacting regions,
i.e. $g_1=0$ and $g_2=g$. 
In this case the  transmission amplitude is a function  
of a dimensionless  parameter $s=\omega/2\pi\rho g$ with the asymptotics
\begin{equation}
t\left(s\right) =\left\{
\begin{array}{l}
-2^{3/4}s^{1/4},\,\,\,\,\,\text{for}\,\,  s \ll 1 \\
\\
1,\,\,\,\,\,\,\,\,  \text{for}\hspace{0.5cm}  s\gg  1. \\
 \end{array}
\right.  \label{eta_b_vs_d}
\end{equation}

\section{Kinetics of 1D Bose fluid:  Thermal current}
\label{s3}

\begin{figure}
\includegraphics[width=0.8\columnwidth,angle=0]{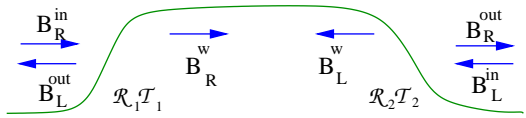}
\caption{Distribution functions for transmitted and reflected plasmons 
inside the wire \vspace{0.4cm}}
\label{plasmons}
\end{figure}

In the preceding section, we have found the plasmon transmission
coefficients at the boundaries between the interacting region and the
reservoirs.  Supplementing these results 
with the boundary conditions on the distribution functions of plasmons
in the reservoirs, we straightforwardly
calculate the distribution functions  
of the right and the left  moving   modes ($B_{R/L}$) inside the wire, see
Fig.~\ref{plasmons}. 

For a wire longer than the thermal wave length of plasmons, the
Fabry-Perot-type plasmon interference can be neglected, and one finds 
\begin{eqnarray}&& 
\label{a32}
B_R^w=\frac{{\cal T}_1}{1-{\cal R}_1{\cal R}_2}B_R^{(0)}+
\frac{{\cal T}_2{\cal R}_1}{1-{\cal R}_1{\cal R}_2}B_L^{(0)}\,,\nonumber  \\&&
B_L^w=\frac{{\cal T}_2}{1-{\cal R}_2{\cal R}_1}B_L^{(0)}+
\frac{{\cal T}_1{\cal R}_2}{1-{\cal R}_2{\cal R}_1}B_R^{(0)}\,,
\end{eqnarray}
where $B_R^{(0)}=\coth (\omega/2T_R)$,  
$B_L^{(0)}=\coth(\omega/2T_L)$, $T_L$ and $T_R$ 
are temperatures of the  left and the right moving collective modes, 
${\cal  T}_1= |t_1|^2$, ${\cal T}_2=|t_2|^2$ are the transmission coefficients
of the left and right barriers, and ${\cal R}_j = 1-{\cal T}_j$ are
the corresponding reflection coefficients.    .

\begin{figure}
\includegraphics[width=0.8\columnwidth,angle=0]{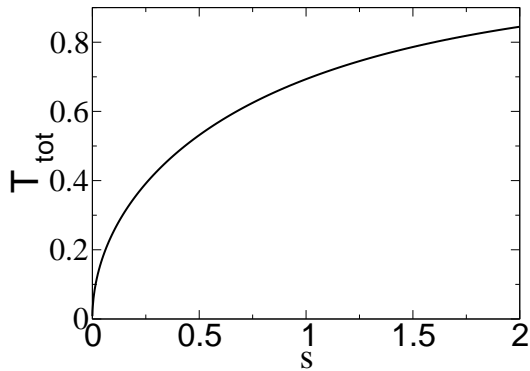}
\caption{Total plasmon transmission for a system of two
  identical sharp barriers as a function of $s = \omega/2\pi \rho g$.}
\label{Ttot}
\end{figure}

Using the plasmon distribution function one can calculate, in
particular, the thermal current  
\begin{eqnarray}
\label{th_a}
I_E=\frac{1}{4\pi}\int_0^\infty d\omega \omega[B_L^w(\omega)-B_R^w(\omega)]\,.
\end{eqnarray}
Substituting Eq.(\ref{a32})  into Eq.(\ref{th_a}), we find
\begin{eqnarray}
\label{IE_1}
I_E=\frac{1}{4\pi}\int_0^\infty d\omega \omega[B_L^0(\omega)-B_R^0(\omega)]
{\cal T}_{\rm tot}(\omega)\,.
\end{eqnarray}
Here ${\cal T}_{\rm tot}$ is the total transmission coefficient 
\begin{eqnarray}
{\cal T}_{\rm tot}(\omega)=\frac{{\cal T}_1(\omega){\cal T}_2(\omega)}{1-{\cal R}_1(\omega){\cal R}_2(\omega)}\,;
\end{eqnarray}
it is shown in Fig.~\ref{Ttot} for the case of two sharp barriers
characterized by the transmission amplitude (\ref{transmission_amplitude}).

\begin{figure}
\includegraphics[width=0.8\columnwidth,angle=0]{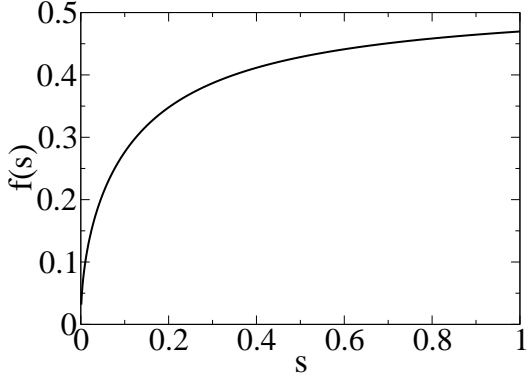}
\caption{Scaling function $f(s)$ governing the dependence of the heat
  current on $s=T/2\pi \rho g$, see Eq.~(\ref{thermal_current})}
\label{f}
\end{figure}

Substituting ${\cal T}_{\rm tot}$ for the case of two sharp barriers
into Eq.~(\ref{IE_1}) and performing the frequency integration, we find
\begin{equation}
\label{IE_2}
I_E = {\cal F}_E(T_L) - {\cal F}_E(T_R)\,,
\end{equation}
where
\begin{equation}
{\cal F}_E (T) =\left\{
\begin{array}{l}
\frac{3}{8\pi}\frac{1}{\sqrt{\rho g}}\xi
\left(\frac{5}{2}\right) T^{5/2}
,\,\,\,\,\,\text{for}\,\,  T \ll 2\pi\rho g \\
\\
\frac{\pi}{12} T^2 \,\,\,\,\,\,\,\, ,  \text{for}\,\, 
T\gg 2\pi\rho g \\
\end{array}
\right.  \label{IE_3}
\end{equation}
In the limit of small temperature difference $\Delta T = T_L - T_R$ 
we have
\begin{eqnarray}
\label{eq_scaling1}
I_E=\frac{\Delta T}{8\pi T^2}\int_0^\infty d\omega 
\frac{\omega^2}{\sinh^2\omega/2T}{\cal T}_{\rm tot}(\omega)\,,
\end{eqnarray}
where $T=\frac{T_L+T_R}{2}$.
Eq.(\ref{eq_scaling1}) can be cast in the scaling form 
\begin{eqnarray}
\label{thermal_current}
I_E=\Delta T T f\left(\frac{T}{2\pi \rho g}\right)\,.
\end{eqnarray}
The function $f$ entering Eq.~(\ref{thermal_current}) can be calculated
numerically and is plotted in Fig.~\ref{f}. 
It has the following asymptotic limits:
\begin{equation}
f\left(s\right) =\left\{
\begin{array}{l}
a s^{1/2},\,\,\,\,\,\text{for}\,\,  s \ll 1 \,, \\
\\
\frac{\pi}{6},\,\,\,\,\,\,\,\,  \text{for}\hspace{0.5cm}  s\gg  1 \,, \\
 \end{array}
\right.  \label{IE_4}
\end{equation}
where  $a \simeq 1.003$.

At relatively high temperatures, $T \gg \rho g$, we reproduce 
the thermal current of non-interacting bosons\cite{Rego}, see
Appendix~\ref{free_bosons}). This result can be considered 
as a thermal-current
counterpart of the Landauer quantization of charge conductance: the
numerical coefficient in the second line of Eqs.~(\ref{IE_3}) and
(\ref{IE_4}) is fully universal in 1D geometry
and does not depend on the form of the
spectrum, interaction strength, carrier statistics (fermions
vs. bosons), etc. The only condition is  the absence of
back-scattering of plasmons.
In the low-temperature regime, we find that the thermal current is suppressed  
due to reflection of the bosons on the boundary between
interacting and  non-interacting regions. This $T^{1/2}$ suppression
of the thermal conductance distinguishes the bosonic setup from its
fermionic counterpart.

\section{Phase coherent  dynamics: Green functions}
\label{s4}

We now proceed with  the analysis of the single-particle  Green
functions of the original bosons,
\begin{eqnarray}&&
\label{GF}
G_B^>(x,\tau)=-i\langle\hat{\Psi}_B(x,t)\hat{\Psi}^\dagger_B(0,0)\rangle\, \,\,, \nonumber   \\&&
G_B^<(x,\tau)=-i\langle\hat{\Psi}^\dagger_B(0,0)\hat{\Psi}_B(x,t)\rangle \,\,\, ,
\end{eqnarray}
that carry information about spectral properties (density of states
and distribution functions) of the system. 
The results  for the  non-interacting case are well known; for completeness 
we present them in Appendix  \ref{free_bosons}.
Our   approach allows us to analyze the Green functions  in
a broad range of parameters, with the only assumption being $T \ll {\rm
  max} (\rho g, \rho^2/2m)$. 
To relate it to well-accepted terminology in the field 
\cite{Gangardt_2003,Gangardt_2009},
the harmonic approximation allows us to describe the system in 
the ``Decoherent Quantum'',  ``Quasi-Condensate'', and ``Tonks-Girardeau''
regimes, and in the strong-interaction LL regime, $\rho g \gg T,
\rho^2/2m$ (which is not realized in the delta-interaction model
considered in Ref.~\onlinecite{Gangardt_2003,Gangardt_2009}), as well as in
crossovers between them.

Using the representation of the boson creation and annihilation
operators in terms of the collective field, Eq.~(\ref{bose_operator}),   
we write the Green function as a correlation function of the harmonic fluid, 
\begin{equation}
\label{a1}
G^\gtrless(x,\tau)=-i\langle T_K
\hat{\rho}_\mp^{1/2}(x,\tau)e^{i\hat{\theta}_\mp(x,\tau)} 
\hat{\rho}_\pm^{1/2}(0,0)e^{-i\hat{\theta}_\pm(0,0)}\rangle\,.
\end{equation}

It is convenient to represent  this correlation function as  
\begin{eqnarray} &&
G^\gtrless(X,\tau)=-i\rho\left(1-i\left(\frac{\partial}{\partial\alpha_1}+
\frac{\partial}{\partial \alpha_2}\right)-\frac{\partial^2}{\partial
\alpha_1 \partial \alpha_2}\right)\nonumber \\&& 
\times \langle e^{i\int (d\omega)dx J^\gtrless_{-\omega,x}
  \Phi^T(\omega,x)} \rangle \,. 
\label{a2}
\end{eqnarray}
Here we have defined a four-component ``source'' vector 
\begin{eqnarray}&&
J^\gtrless_{1,-\omega,x}=\frac{-1}{2\pi\rho\sqrt{2}}\frac{\partial}{\partial x}
\bigg[e^{-i\omega\tau}\alpha_1\delta(x-X)+\alpha_2\delta(x)\bigg],
\nonumber \\&&
J^\gtrless_{2,-\omega,x}=\frac{1}{\sqrt{2}}\bigg[e^{-i\omega\tau}\delta(x-X)-\delta(x) \bigg]\,, \nonumber \\&&
J^\gtrless_{3,-\omega,x}=\frac{\pm 1}{2\pi\rho \sqrt{2}}\frac{\partial}{\partial x}
\bigg[e^{-i\omega\tau}\alpha_1\delta(x-X)-\alpha_2\delta(x)\bigg],\nonumber 
\\&&
J^\gtrless_{4,-\omega,x}=\mp\frac{1}{\sqrt{2}}
\bigg[e^{-i\omega\tau}\delta(x-X)+\delta(x) \bigg]\,,
\end{eqnarray}
and it is understood that one should set the sources $\alpha_{1,2}=0$
after the derivatives in Eq.~(\ref{a2}) have been evaluated.

Since the action (\ref{a4}) is Gaussian, the functional integration over 
fields $\theta$ and $\phi$ can be easily performed, yielding
\begin{eqnarray}&&
\label{a23}
G^\gtrless(X,\tau)=-i\rho\left(1-i\left(\frac{\partial}{\partial\alpha_1}+
\frac{\partial}{\partial \alpha_2}\right)-\frac{\partial^2}{\partial
\alpha_1 \partial \alpha_2}\right)\nonumber  \\&& 
\exp\left(-\frac{i}{4}\int (d\omega)dx_1dx_2
  J^\gtrless_{-\omega,x_1}D_{\omega,x_1,x_2} 
J^{\gtrless T}_{\omega,x_2}\right) \,. \nonumber
\end{eqnarray} 
Thus, the problem of calculation of the Green functions has been reduced 
to the calculation of the bosonic propagator 
\begin{eqnarray}
D=\begin{pmatrix}
D^K & D^{r}\\
D^{a} & 0
\end{pmatrix}.
\label{a5}
\end{eqnarray}

We now focus on the case of coinciding spatial points ($X=X'$)
deeply inside   
the interacting part of the wire (for  $X \neq X'$ see Appendix \ref{GF_int}). 
Expanding the bosonic density operator up to second order in
$\phi$,  and calculating Gaussian functional integral over the bosonic
fields (see Appendix \ref{GF_int} for technical details), we obtain
\begin{eqnarray}
\label{a29}
G^\gtrless(\tau)
=-i\rho\bigg(1-\Phi_1^\gtrless(\tau)\bigg)e^{-\Phi_2^\gtrless(\tau)}\,. 
\end{eqnarray}
Here we have defined the pre-exponential factor
\begin{eqnarray}&&
\label{a30}
\Phi^\gtrless_1(\tau)=\frac{m}{2\pi\rho}\int_0^\infty
\frac{d\omega}{4\rho gq  +q^3/m} 
\bigg[
(B_R^w+B_L^w)\\&&
\times\left(
i\omega \sin\omega\tau
-\frac{q^2}{4m}\cos\omega\tau
\right)\pm
\frac{iq^2}{2m}\sin\omega\tau
\mp 2\omega\cos\omega\tau
\bigg)\nonumber\,,
\end{eqnarray}
and the exponential factor
\begin{eqnarray}&&
\label{a31}
\Phi^\gtrless_2(\tau)=\frac{m}{2\rho}\int_0^\infty \frac{d\omega}{2\pi}
e^{-\omega/\Lambda}
\frac{g+q^2/8m\rho}{q( g+q^2/4m\rho)}\nonumber \\&&
\times \bigg((B_R^w+B_L^w)
(1-\cos\omega\tau)\pm 2i\sin \omega\tau
\bigg)\,.
\end{eqnarray}
In these formulas $q$ should be understood as related to $\omega$ via the
dispersion law (\ref{q}). 

It is instructive to compare our results for the Green
functions of a  bosonic system in a partial non-equilibrium state,
with their fermionic counterparts \cite{gutman09}. It is seen that
the bosonic results differ in two respects: (i) the appearance of a
pre-exponential factor and (ii) a more complicated form of the
exponential factor reflecting the non-linear character of the spectrum
of collective excitations. 

In fact, Eq.~(\ref{a29}) interpolates between the standard-LL results
(applicable in the limit of strongly interacting bosons) and Eq.(\ref{a300}) 
which is valid for  free bosons. 
The characteristic energy scale for the crossover between these two regimes  
is  set  by the interaction ($\omega_0=\rho g$). For energies well below 
$\omega_0$,  the model behaves as a standard LL system 
(with a linear plasmonic spectrum). 
In this case the pre-exponential factor $\Phi_1$ is small
(of the order of $T/\omega_0 \ll 1$), and therefore can be neglected.
The exponential factor $\Phi_2$ in this limit turns into the LL
result of Ref.~\onlinecite{gutman09} (which, of course, reduces to the
conventional LL formula in the equilibrium case). In  the high-energy limit 
$({\rm max}\{\tau^{-1}, T\} \gg \omega_0)$ the  
spectrum of excitations is quadratic, and one recovers the 
free-boson results, see Appendix \ref{free_bosons}. 

\section{Summary}
\label{s5}

In this work, we have studied a system of  bosonic atoms in   
a  Landauer setup,  subject  to  
temperature and chemical potential difference.
We have developed a non-equilibrium bosonization approach that 
describes the system within the harmonic approximation.
This approach ignores the interaction between different collective modes
but  takes into account the non-linear dispersion of their spectrum.

We have studied the plasmon propagation in a two-terminal setup formed
by two non-interacting reservoirs connected by an interacting 1D
``wire''.  The plasmon back-scattering is controlled by the
dimensionless parameter $s \sim \omega/\rho g$, where $\omega$ is the plasmon
frequency (whose characteristic value is set by the respective  
temperatures of the reservoirs), 
$\rho$ is the density of bosons, and $g$ the interaction
strength. At large $s$, back-scattering is suppressed, and the  
thermal current acquires a universal value (that depends neither on
the spectrum of the  original particles nor on  their statistics). In
the low-temperature regime, reflection of plasmons is strong, and 
the thermal current exhibits  a $T^{1/2}$ suppression compared to the
universal value. As another application of this formalism, 
we have calculated  the bosonic single particle Green function in the
same non-equilibrium setup. The result interpolates between
the conventional (linear spectrum) Luttinger-liquid and free-boson limits. 

The approach developed in this work can be used to study other
properties of the system, e.g., higher correlation functions or
non-steady-state characteristics. A more ambitious perspective will be  
development of a general non-equilibrium bosonized theory of non-linear LL
(formed by interacting fermionic or bosonic particles) 
including both non-linear spectral dispersion of plasmon modes and
their coupling.

\section{ Acknowledgments}

We thank N. Davidson,  L. Khaykovich, and  J. Schmiedmayer
for useful discussions.
Financial support by German-Israeli Foundation,the  Israel Science Foundation, 
and the Minerva Center is gratefully acknowledged.

\appendix
\section{Transport properties of free bosons}
\label{free_bosons}

In this Appendix we summarize basic properties of non-interacting bosons, 
using the original description.   

\subsection{Landauer approach:  Particle  current}

For the case of non-interacting bosons the particle current can be
straightforwardly found within the Landauer approach: 
\begin{equation}
I=\int_0^\infty d\epsilon
\nu(\epsilon)v(\epsilon)[N_L(\epsilon)-N_R(\epsilon)]\,, 
\end{equation}
where $\nu$ is the density of states, $v$ the velocity, and $N_j$ the
distribution function in the corresponding reservoir.  
Using the relation $\nu(\epsilon) v(\epsilon)=1/2\pi$, one finds the relation
\begin{equation}
I=\int_0^\infty \frac{d\epsilon}{2\pi} [N_L(\epsilon)-N_R(\epsilon)]\,.
\end{equation}
Assuming Bose distributions with the same temperature $T$ and two
different chemical potentials $\mu_L$, $\mu_R$ and performing the
integration, we obtain
\begin{eqnarray}
I=-\frac{T}{2\pi}\ln\left(\frac{1-e^{\mu_L/T}}{1-e^{\mu_R/T}}\right).
\end{eqnarray}
We note that the particle current is determined by the occupation
numbers in both reservoirs 
at the bottom of the  band (at $\epsilon=0$). 
In the linear response regime, expanding the current in the difference
of chemical potentials, we find
\begin{equation}
I=\frac{1}{2\pi}(\mu_L-\mu_R)N_B(\epsilon=0)\,.
\end{equation}  
To relate the occupation number at the bottom of the band to
macroscopic parameters of the problem we compute the  particle density 
\begin{equation}
\int_0^\infty d\epsilon \nu(\epsilon)N_B(\epsilon)=\rho\,.
\end{equation}
For $|\mu| \ll T$ one finds 
\begin{equation}
\label{chemical_potential}
\rho=T\sqrt{\frac{m}{|\mu|}}\,,
\end{equation}
thus leading to $N_B(\epsilon=0)=\rho^2/mT$
and 
\begin{equation}
\label{current_dc}
I=\frac{1}{2\pi}\frac{\rho^2}{mT}(\mu_L-\mu_R)\,.
\end{equation}
We note that the particle current of bosons is enhanced by a large parameter 
$\rho^2/mT$ as compared to the particle current of fermions subjected 
to the same difference of chemical potentials. 
The large conductance is reminiscent of  superfluidity of Bose
condensate in higher dimensions.  
 
On a more formal level, the appearance of the ``Fermi energy''
$\rho^2/m$ entering the large factor $\rho^2/mT$ indicates 
that the particle current  can not be calculated  
within the harmonic approximation. In the interacting case its
calculation would require the use of a full non-linear  
hydrodynamic theory, see Appendix \ref{appendix:GF} below.

\subsection{Thermal current}

The thermal current is given by 
\begin{equation}
I_E=\int_0^\infty \frac{d\epsilon}{2\pi} \epsilon [N_L(\epsilon)-N_R(\epsilon)]
\end{equation}
Performing the integration one finds
\begin{equation}
I_E=\frac{\pi}{12}(T_L^2-T_R^2)\,.
\end{equation}
We note that the heat current in the system of  non-interacting bosons 
coincides with the heat current of 1D free fermions\cite{Fazio}.
Moreover, this result is universal and does not depend on the shape of 
single particle spectrum of elementary carriers, either fermions or bosons\cite{Rego}.
This universality  survives the adiabatic switching of interaction
(again for both fermions and bosons), when back-scattering of plasmons
is negligible,  but is ultimately violated in the generic
case [see, in particular, Eq.(\ref{thermal_current})].

\section{Green functions, $G^\gtrless$}
\label{appendix:GF}

In this Appendix, we present details of the calculation of the bosonic
Green function. We begin by considering the case of non-interacting
bosons, first in the original formulation and then in the within the
bosonization framework. Then we calculate the Green function in the
interacting case.

\subsection{Non-interacting bosons: Original description }
\label{GF_appendix}

We consider  the non-interacting 1D bosons 
and  calculate the Green functions for coinciding spatial points,
$X=X'$ (the case of $X \neq X'$ treated similarly).   
Equations (\ref{GF}) yield 
\begin{equation}
\label{a300}
G^\gtrless_B(\tau)=-\frac{i}{2\pi}\sqrt{\frac{m}{2}}\int_0^\infty
\frac{d\omega}{\sqrt{\omega}}e^{-i\omega\tau} 
\left(\coth\frac{\omega-\mu}{2T}\pm 1\right)\,.
\end{equation}
In the energy representation  it reproduces well known results  
\begin{eqnarray}&&
\label{a301}
G^<(\omega)=-2\pi i \nu(\omega) N_B(\omega-\mu)\,,\\&&
G^>(\omega)=-2\pi i\nu(\omega)[1+N_B(\omega-\mu)]\,,
\end{eqnarray}
where $N_B$ is the Bose distribution function  and 
\begin{equation}
\label{a302}
\nu(\omega)=\frac{\theta(\omega)}{\pi}\sqrt{\frac{m}{2\omega}}\,
\end{equation}
is the density of states of non-interacting bosons.
To compare with the bosonized theory, we consider the high-density ($\mu \rightarrow 0$) limit.
Next, we show how these results can be derived using the bosonization approach.

\subsection{Non-interacting bosons: Bosonized description}

The Green function within the bosonization framework is given by
Eq.~(\ref{a29}). 
Substituting  the spectrum of free bosons in Eq.~(\ref{a30}), we get 
\begin{eqnarray}&&
\Phi_1^\gtrless(\tau)= \frac{1}{2\pi \rho}\sqrt{\frac{m}{2}}
\int_0^\infty\frac{d\omega}{\sqrt{\omega}}\bigg[
B_\omega(i\sin\omega\tau-\frac{1}{2}\cos\omega\tau)\nonumber \\&&
\pm\left(\frac{i}{2}\sin\omega\tau-\cos\omega\tau
\right)
\bigg]\,.
\label{a33}
\end{eqnarray}
Similarly, Eq.(\ref{a31}) yields for the function in the exponent, 
\begin{eqnarray}&&
\Phi_2^\gtrless(\tau)= \frac{1}{4\pi \rho}\sqrt{\frac{m}{2}}
\int_0^\infty\frac{d\omega}{\sqrt{\omega}}\bigg[
(1-\cos\omega\tau)B_\omega\nonumber \\&&\pm i\sin\omega\tau
\bigg]
\label{a33a}
\end{eqnarray}
Let us emphasize that the integral over frequency in Eq.(\ref{a33}) converges
(as opposed to the logarithmically divergent integrals in the LL case).
The  resulting function $\Phi_2$ is actually small.
Therefore, one can expand the exponent in Eq.(\ref{a29}).
Combining all terms together, one finds
\begin{equation}
G^\gtrless(\tau)\simeq
i\rho\left(-1+\Phi_1^\gtrless(\tau)+\Phi_2^\gtrless(\tau)\right)\,. 
\label{a34}
\end{equation}
To compare this with the exact result, we take Eq.~(\ref{a300}),
subtract its value at $\tau=0$ (density), and then consider the limit
$\mu\to 0$. The result is in full agreement with Eqs.~(\ref{a34}),
(\ref{a33}), and (\ref{a33a}).

\subsection{Interacting bosons,  Bosonized description}
\label{GF_int}
We now consider the case  both coordinates $X$ and $X'$
are located  inside the interacting part of the system.
In this case the Green function  depends on the $x=X-X'$.
For the problem of the sharp barrier one finds
\begin{eqnarray}
G^\gtrless(\tau,x)=-i\rho\bigg(1-\Phi_1^\gtrless(\tau,x)\bigg)e^{-\Phi_2^\gtrless(\tau,x)+Ix}\,.
\end{eqnarray}
Here the pre-exponential 
\begin{eqnarray}&&
\Phi_1^\gtrless(\tau,x)=
\frac{m}{2\pi\rho}\int_0^\infty
\frac{d\omega}{4\rho gq  +q^3/m} 
\bigg[
B_R^w
\bigg(i\omega\sin\xi_R \nonumber  \\&&-\frac{q^2}{4m}\cos\xi_R 
\bigg)\pm\frac{iq^2}{4m}\sin\xi_R\mp\omega\cos\xi_R
+ (R\leftrightarrow L)\bigg],\nonumber
\end{eqnarray}
and exponential factor
\begin{eqnarray}&&
\Phi_2^\gtrless(\tau)=\frac{m}{2\rho}\int_0^\infty \frac{d\omega}{2\pi}
\frac{g+q^2/8m\rho}{q(g+q^2/4m\rho)}\nonumber \\&&
\bigg(B_R^w
(1-\cos\xi_R)\pm i\sin\xi_R +(R\leftrightarrow L)
\bigg)\,,
\end{eqnarray}
where $\xi_{R/L}=\omega \tau\mp qx$ and 
$I$ is  an average value of the particle current flowing through the system,
see Appendix  \ref{non_linear}.
It is given by the mean value of $\langle \partial_x\theta \rangle$, 
which is determined by a non-linear hydrodynamic equation.

We note that calculating the correlation function of the bosonic fields
we neglected the modulation of the mean density of bosons.
It is possible to generalize the harmonic approximation and to 
allow the modulation of the mean density in space.
We now present general formulas applicable in the case of spatially varying mean density.

To calculate the correlation functions of bosonic fields, 
one needs to find the inverse of the operator 
\begin{equation}
D^{-1}D=\hat{1}\,.
\end{equation}
Employing Eq. (\ref{Polarization_operator2}), we find
that components of the matrix correlation function $D$ satisfy the following equations:
\begin{eqnarray}&&
\hat{L}D^r_{\phi\phi}(\omega,x,x')=-\frac{2\pi^2\rho}{m}\hat{1}\,,
\\&&
\hat{L} D^r_{\phi\theta}(\omega,x,x')=2\pi i \omega\partial_x^{-1}\hat{1} \,,
\nonumber \\&&
\hat{L}
D^r_{\theta\phi}(\omega,x,x')=2\pi i \omega\hat{1}\,,
\nonumber \\&&
\hat{L} \partial_x D^r_{\theta\theta}(\omega,x,x')=
(2\pi)^2\hat{K}^{-1}(x)\partial_x^{-1}\hat{1}, \nonumber 
\end{eqnarray}
where we have defined an operator 
\begin{equation}
\hat{L}\equiv \omega^2-\frac{2\pi^2\rho}{m}\hat{K}^{-1}(x)\,.
\end{equation}
In order to find the components of the correlation function  $D$,
we construct the scattering 
states $\chi_{q,\eta}(x)$, characterized by the momentum $q$ and 
index $\eta$ that labels the reservoir from which the state was ``emitted''.
The scattering state wave function satisfy the following equation
\begin{equation}
\frac{2\pi^2\rho}{m}\hat{K}^{-1}
\chi^\eta_q(x)=\omega^2_q\chi^\eta_{q}(x)\,.
\end{equation}

In terms  of the scattering states, the correlation functions of the bosonic fields can be written as follows:
\begin{eqnarray}&&
\label{a22}
D_{\phi\phi}^{r/a}(\omega,x,x')=
\frac{2\pi^2\rho}{m}
{\cal D}^{r/a},\\&&
D_{\phi\theta}^{r/a}(\omega,x,x')=-2\pi i \omega \partial_{x'}^{-1}\frac{2\pi^2\rho}{m}{\cal D}^{r/a}, \nonumber \\&&
D_{\theta\phi}^{r/a}(\omega,x,x')=
2\pi i \omega \partial_{x}^{-1}
\frac{2\pi^2\rho}{m}
{\cal D}^{r/a}, \nonumber \\&&
D_{\theta\theta}^{r/a}(\omega,x,x')= 
-(2\pi)^2\hat{K}^{-1}(x')\partial_x^{-1}\partial_{x'}^{-1}
{\cal D}^{r/a}. \nonumber 
\end{eqnarray}
Here we defined 
\begin{eqnarray}
\label{a25}
{\cal D}^{r/a}=
\sum_{q>0,\eta}
\frac{\chi^\eta_{q}(x)\chi^{\eta *}_{q}(x')}{\omega_\pm^2-\omega^2_q}.
\end{eqnarray}
The Keldysh component can be constructed from the retarded and advanced one, 
by imposing the ``partial equilibrium'' condition in each direction of plasmon propagation,
\begin{equation}
\label{a24}
{\cal D}^K=-2\pi i
\sum_{q>0,\eta}\chi^\eta_{q}(x)\chi^{*\eta}_{q}(x')B_\eta(\omega)\delta(\omega^2-\omega_q^2),
\end{equation}
and similarly for other components.

\subsection{Mean density current for interacting bosons}
\label{non_linear}
To find the particle current one 
needs to calculate the expectation value of the  operator
\begin{equation}
I=\frac{\rho}{m}\partial_x\theta(x)\,.
\end{equation}
In the  presence of a weak external potential $U(x,t)$ 
the linear response theory predicts
\begin{equation}
\label{conductance}
\langle I(x,t) \rangle= 
\partial_t\int dx'dt'D^r_{\phi\phi}(x,t;x',t')\partial_xU(x',t')\,.
\end{equation}
We now analyze this expression in different limits. 
If  the interaction between bosons 
in the leads $g_r$ is finite 
(we now relax the assumption $g_r=0$ we used throughout the manuscript),  
it sets the energy scale $\rho g_{\rm r}$. 
For energies below this scale the spectrum of collective modes is linear,
and  one restores  the known LL result \cite{Kane,maslov-lectures}
\begin{equation}
\langle I\rangle=K_r V/h\,.
\end{equation}
Here $K_{r}=\sqrt{\pi^2\rho/2mg_r}$ is the LL  parameter in the leads, 
and $V$ is a difference between  the chemical potentials in  the leads.
In the absence of interactions in the leads,  the value of current in this case
is unaffected by the interaction inside the system, as in the fermionic case\cite{Maslov,Safi,Ponomarenko,Safi97,Oreg,Thomale}.

For the case of free  bosons ($g_r=g=0$)
Eq.(\ref{conductance}) yields
\begin{equation}
\langle I(\omega)\rangle=\frac{\rho}{2m}\sqrt{\frac{m}{2\omega}}V\,,
\end{equation}
implying that  the ``conductance'' diverges in 
the d.c. limit ($\omega\rightarrow 0$).
This divergence signals that  the harmonic approximation
is not a suitable framework to calculate the particle current. 
Indeed,  while developing  the harmonic approximation 
we took the $\mu \rightarrow 0$ limit, 
assuming that characteristic energies of the collective excitations are 
much  greater than the chemical potential. 
While this assumption is valid for the thermal current and 
the single-particle Green function it does not hold for the particle current 
of the free bosons.  
To cure this problem, one should restore the low energy cut-off, 
replacing  $\omega$   with  $\mu$. 
Doing so and using Eq.(\ref{chemical_potential}), 
one recovers the conductance of non-interacting bosons, Eq.(\ref{current_dc}).

While the limits discussed above give us some idea about the particle current, 
this problem remains  to be solved   
for bosons that do interact in the wire ($g \neq 0$),  
but do not interact in the leads ($g_r=0$). 
In particular,  it remains to be seen whether the value of particle 
current in this case is affected by interaction inside the system.
To answer this question, one has to go beyond the harmonic approximation 
and to resort to  the full (non-linear) hydrodynamic description.

\end{document}